\documentclass[aps,prb,twocolumn,amsmath,amssymb]{revtex4}
\usepackage{times}
\usepackage{graphicx}
\usepackage{amsfonts}
\usepackage{amsmath, amsthm, amssymb}
\usepackage{dsfont}

\newcommand{\e}[1]{\mathrm{e}^{#1}}

\newcommand{\eg}{\textit{e.g. }}
\newcommand{\etal}{\emph{et al.}}
\def\i{\mathrm{i}}

\begin{document}

\title{Reversal and Termination of Current-Induced Domain Wall Motion \\
via Magnonic Spin-Transfer Torque}

\author{Jacob Linder}

\affiliation{Department of Physics, Norwegian University of
Science and Technology, N-7491 Trondheim, Norway}
 
\begin{abstract}
We investigate the domain wall dynamics of a ferromagnetic wire under the combined influence of a spin-polarized current and magnonic spin-transfer torque generated by an external field, taking also into account Rashba spin-orbit coupling interactions. It is demonstrated that current-induced motion of the domain wall may be completely reversed in an oscillatory fashion by applying a magnonic spin-transfer torque as long as the spin-wave velocity is sufficiently high. Moreover, we show that the motion of the domain wall may be fully terminated by means of the generation of spin-waves, suggesting the possibility to pin the domain-walls to predetermined locations. We also discuss how strong spin-orbit interactions modify these results.
\end{abstract}
 
\date{\today}

\maketitle

\section{Introduction}

The topic of domain-wall motion in magnetic nanowires has in the last years become subject to much interest \cite{schryer_apl_74, atkinson_natmat_03, beach_prl_06, slonberger, zhang_prl_02, zhang_prl_04, grollier_apl_03, hayashi_prl_06, parkin_science_08, tserkovnyak_08, li_prl_08, khvalkovskiy_prl_09}. This is in large part due to the prospect of transferring and saving information via manipulation of domains that distinguish opposite directions of magnetization. Finding ways to achieve such domain-wall motion with minimal energy losses due to Joule heating from resistance and at the same time retain well-defined control over the dynamics is crucial with respect to possible technological applications \cite{tretiakov_prl_10}.

It is known that domain-wall motion may be induced by either injecting electric currents \cite{slonberger, zhang_prl_02, zhang_prl_04, grollier_apl_03, hayashi_prl_06, parkin_science_08, tserkovnyak_08, li_prl_08, khvalkovskiy_prl_09} into nanowires or utilizing static magnetic fields \cite{schryer_apl_74, atkinson_natmat_03, beach_prl_06}. However, a second opportunity was very recently revealed: namely, moving the domain-wall structure via magnonic spin-transfer torque \cite{han_apl_09, jamali_apl_10, yan_prl_11}. The basic idea behind this mechanism is that the magnon spin is flipped when passing through a domain wall indicating that spin angular momentum has been absorbed by the domain wall, thus causing it to move in the opposite direction of the spin-wave. 

Interestingly, the properties of spin-transfer torque phenomena should be modified in the presence of spin-orbit interactions \cite{obata_prb_08, hals_prl_09, haney_prl_10, manchon_prb, moore_apl_08, kim_arxiv_11}. Such interactions may originate from intrinsic properties of the material at hand or the structural inversion asymmetry present at surfaces. A series of recent experiments \cite{miron} have reported anomalous behavior of domain-wall motion due to current-bias which could possibly be attributed to Rashba spin-orbit effects in the sample. This points to the necessity of obtaining a fuller understanding of the role played by this type of interactions.

Motivated by the recent progress in this field, we pose the following question in this paper: what happens under the combined effect of current-induced dynamics and magnonic spin-transfer torque? Moreover, how will these phenomena be modified in the presence of spin-orbit interactions? We will show that the interplay between the torques induced by itinerant spin-polarized electrons, spin-$\frac{1}{2}$ particles, and magnons, spin-1 excitations, offers the possibility of \textit{reversed domain-wall motion in an oscillatory fashion}. More specifically, we shall demonstrate that the angular momentum transfer from spin-waves is able to completely reverse the motion of a domain-wall even after it has been accelerated by an electron-current. In addition to this, we demonstrate that the domain-wall motion may be fully terminated at predetermined locations due to the behavior of the magnonic spin-torque. The predicted  effects do not rely on the presence of spin-orbit coupling, but we nevertheless study how the latter, if present, influences the domain-wall dynamics. The domain wall reversal and termination persists in the presence of weak spin-orbit interactions, whereas the dynamics display a very rich, yet complicated, behavior for stronger magnitude of the spin-orbit coupling. These results are suggestive in terms of finding novel ways to manipulate domain-wall motion via an interplay of conventional current-induced dynamics with magnonic spin-transfer torque. 

\begin{figure}[hbt!]
\centering
\resizebox{0.46\textwidth}{!}{
\includegraphics{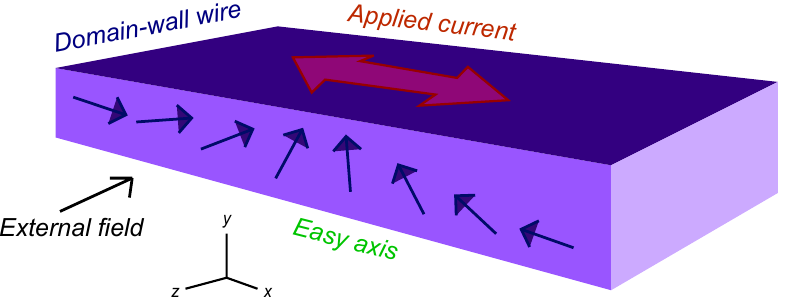}}
\caption{(Color online) The system under consideration: a ferromagnetic wire with a head-to-head domain wall along the easy axis. A spin-polarized current injected through the wire moves the domain-wall due to angular momentum transfer. We also include the possibility of a locally applied external magnetic field. }
\label{fig:model} 
\end{figure}

\begin{figure*}[t!]
\centering
\resizebox{1.0\textwidth}{!}{
\includegraphics{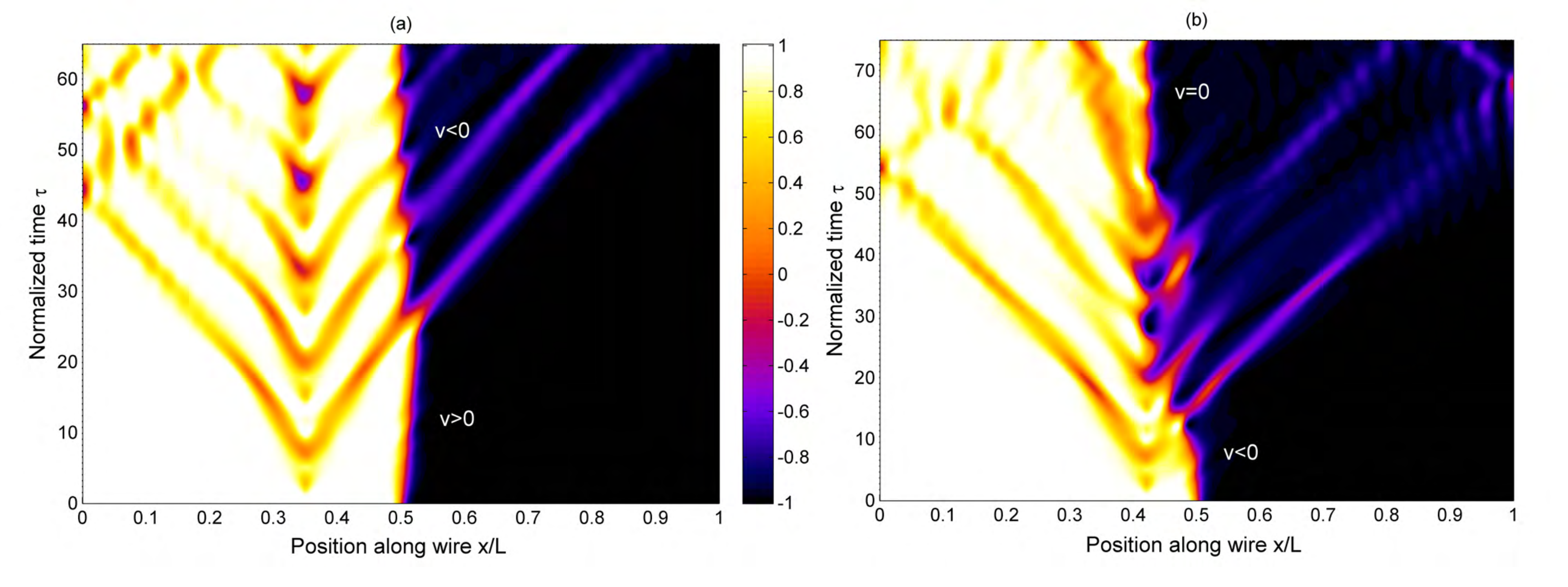}}
\caption{(Color online) Contour-plot of $m_x$ as a function of time and position along the ferromagnetic wire. We have used the parameters $\alpha=0.01, \beta=0.005$, $A=8\times10^{-5}$ with $K_x=-0.5$, $\mathcal{H}_\text{ext,0}=2$, $\omega=0.75$ and current $|J|=1\times10^{-3}$ (corresponding parameters with real units are given in \cite{para}) The spin-orbit interaction parameter is set to zero for simplicity, but the qualitative behavior is unchanged up to $|J_\alpha|\sim0.01$. (a): The domain-wall driven motion due to current-injection and magnon propagation have opposite velocity. (b): Equal velocities.}
\label{fig:without} 
\end{figure*}

\section{Theory}

We consider a model as shown in Fig. \ref{fig:model}: a magnetic wire of length $L$ with a head-to-head domain-wall. Current may be injected into the system and the easy axis is chosen along the extension of the wire $(\hat{x})$. The dynamics of the local magnetic order parameter is governed by the Landau-Liftshitz-Gilbert (LLG) equation, which for our purposes may be cast in the following dimensionless form \cite{thiaville_epl_05, kim_arxiv_11}:
\begin{align}\label{eq:llg}
\partial_\tau \mathbf{m} &= -\mathbf{m}\times[\mathcal{H}_\text{eff} + J_\alpha\hat{y} + \beta J_\alpha(m_z\hat{x}-m_x\hat{z})] \notag\\ &+ \alpha\mathbf{m}\times\partial_\tau\mathbf{m} +J\partial_\zeta\mathbf{m} -\beta J\mathbf{m}\times\partial_\zeta\mathbf{m}.
\end{align}
Here, $\tau=t\mu_0\gamma M_S$ and $\zeta=x/L$ are the normalized time- and spatial-coordinates, whereas $\gamma$ is the gyromagnetic ratio and:
\begin{align}
J = \frac{\mu_BPj_e}{e\gamma M_s^2L(1+\beta^2)},\; J_\alpha = \frac{\alpha_R m_ePj_e}{\hbar e M_s^2(1+\beta^2)}.
\end{align}
The coefficients $J$ and $J_\alpha$ are thus measures for the current-induced spin-transfer torque and the effective field originating from the Rashba spin-orbit interaction, respectively. Above, $P$ is the spin-polarization of the injected current, $j_e$ is the current density, $\alpha$ is the Gilbert damping constant, $\beta$ is the non-adiabaticity parameter, $e$ is the electron charge, $\alpha_R$ represents the spin-orbit interaction strength, while $\mu_B$ is the Bohr magneton. The term $\mathcal{H}_\text{eff}$ contains the contribution to the effective field from anisotropy, exchange stiffness, and external applications:
\begin{align}
\mathcal{H}_\text{eff} = K_xm_x\hat{x} + A\frac{\partial^2\mathbf{m}}{\partial\zeta^2} + \mathcal{H}_\text{ext}(t).
\end{align}
We have introduced the normalized anisotropy coefficient $K_x=H_x/(\mu_0 M_S)$, stiffness constant $A=2A_\text{ex}/(\mu_0M_s^2L^2)$, and  $\mathcal{H}_\text{ext}=H_\text{ext}/(\mu_0M_s)$. Here, $M_s$ is the saturation magnetization and $\mu_0$ is the magnetic permeability. It is noticed that the spin-orbit interaction in general induces two types of torques, both included in Eq. (\ref{eq:llg}): a standard field-like one in addition to a Slonczewski-like torque, the latter one proportional to $\beta$. In fact, such a torque was proposed in order to explain magnetization reversal by an in-plane current in \cite{miron}, indicating that the field-like description of the spin-orbit torque would not suffice. 

Obtaining a general analytical solution of Eq. (\ref{eq:llg}) is a formidable task. The Walker prescription for a domain wall does not hold in general when including both magnon-induced torque, current-induced torque, and spin-orbit coupling, and is solvable only some limiting cases which narrows down the regime of validity considerably. For this reason, we proceed to solve Eq. (\ref{eq:llg}) in its full form numerically without any simplifying assumptions. The coupled equations for the components $m_j$ of the magnetization are supplemented with von Neumann boundary conditions in order to obtain the local time-resolved magnetic order parameter $\mathbf{m}(x,t)$. 

\section{Results and Discussion}
We are now interested in considering the interplay between conventional current-induced domain wall motion and the dynamics generated by magnonic spin-transfer torque absorption. To this end, we consider the scenario shown in Fig. \ref{fig:model}: a spin-polarized current is injected into the ferromagnetic wire and causes the domain-wall to move along the $+\hat{x}$-direction. A spin-wave of frequency $\omega$ is generated by applying an external time-dependent field $\mathcal{H}_\text{ext,0}\sin(\omega \tau)\hat{z}$ locally modelled with a Gaussian function of small width. The position chosen for the local field ensures that there is no influence of spin-waves reflected at the ends of the magnetic nanowire within the considered time-domain. The domain-wall center is initially situated in the middle of the wire ($x=0.50$). The resulting dynamics are shown in Fig. \ref{fig:without}(a), where a contour-plot is shown of $m_x$ in the $(x,t)$-plane. At $t=0$, the magnetization changes sign at $x=0.50$ due to the domain-wall structure. The center of the domain-wall moves right due to the current injection at the same time as a spin-wave starts to propagate in the same direction. As the velocity of the spin-wave $v_\text{spin}$ exceeds the domain-wall velocity $v_\text{DW}$, the spin-wave eventually impacts the center of the domain-wall and deposits its spin-transfer torque. Remarkably, the domain-wall now reverses its direction and starts to propagate in the opposite direction in spite of the persistent spin-polarized current. 

\begin{figure*}[t!]
\centering
\resizebox{1.0\textwidth}{!}{
\includegraphics{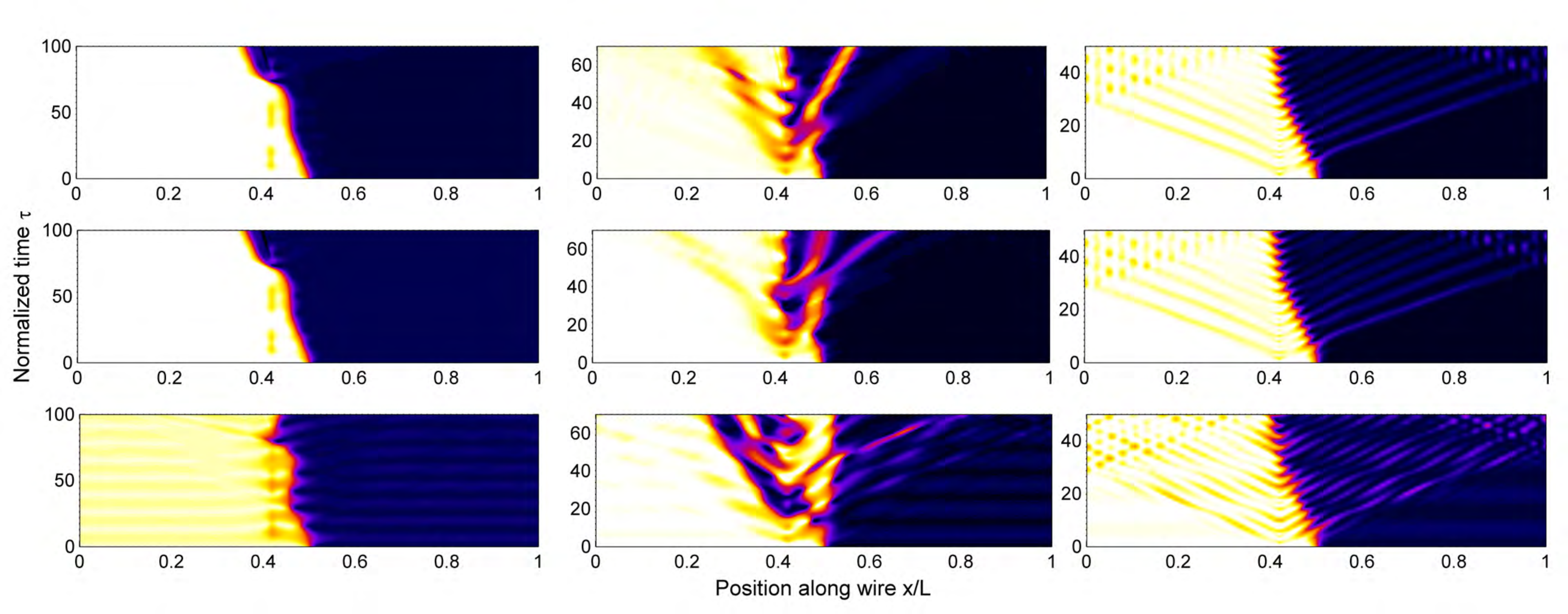}}
\caption{(Color online) Contour-plot of $m_x$ as a function of time and position along the ferromagnetic wire. The same parameters as in Fig. \ref{fig:without} have been used except for the frequency and spin-orbit strength: the columns correspond to $\omega=0.1, 0.5, 1.5$ while the rows correspond to $J_\alpha=0,0.01,0.1$. The current is applied in \textit{same direction} as the spin-wave driven domain-wall motion.}
\label{fig:with1} 
\end{figure*}

The reversal of the domain-wall motion is seen to occur in a non-monotonous way. The turning points in the nearly oscillatory motion are seen to coincide with the maxima of the spin-wave amplitudes reaching the center of the domain-wall. This suggests that there is a persistent competition between the electric-current and magnonic spin-transfer torque: the absorption of angular momentum from the magnons reverses the domain-wall motion each time a spin-wave hits the center, whereas the electric current attempts to move it in the opposite direction. The net motion, however, is still reversed by the magnon-induced torque. 

We now contrast this finding with a scenario where the current-induced and magnonic torque conspire to move the domain-wall in the same direction. This is shown in Fig. \ref{fig:without}(b). A previous work considering this particular situation \cite{jamali_apl_10} found that the spin-waves could enhance the domain-wall velocity, as is also seen in Fig. \ref{fig:without}(b): the domain-wall displays a small yet finite acceleration due to the angular momentum transfer from the magnons (near $\tau\simeq5$ in the plot). However, an additional effect may be inferred from this plot. When the domain-wall center reaches the spin-wave source, i.e. the position of the locally applied external field, it is arrested. The domain-wall structure remains well-defined, with its center now pinned \textit{in spite} of the spin-polarized current injection. Although the locally applied external field poses an experimental challenge, there exists present-day techniques to accomplish this aim in magnetic nanowires \cite{kim_jjap_04, bergmann_apl_09} based on \eg generating current pulses through levels of copper metal lines in order to generate local fields along the nanowire. This effect would make possible controllable movement of the domain-wall to pre-determined locations. 

\begin{figure*}[t!]
\centering
\resizebox{1.0\textwidth}{!}{
\includegraphics{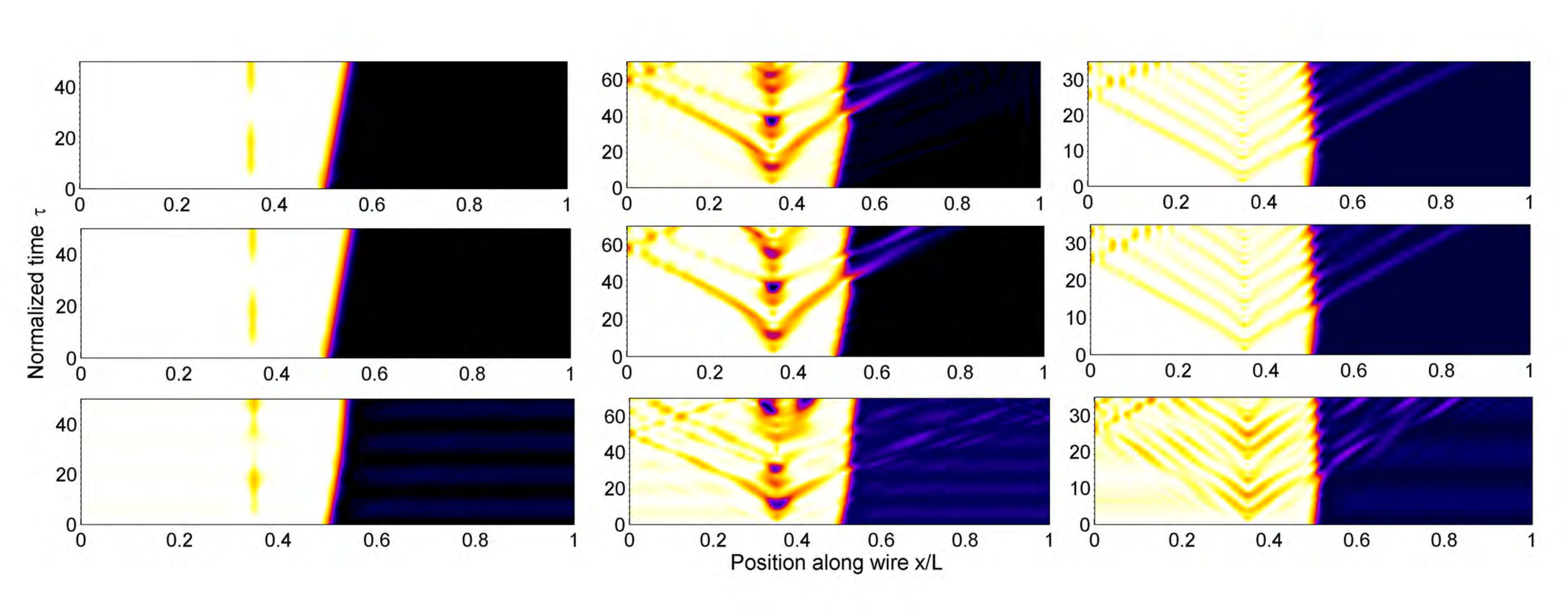}}
\caption{(Color online) Contour-plot of $m_x$ as a function of time and position along the ferromagnetic wire. The same parameters as in Fig. \ref{fig:without} have been used except for the frequency and spin-orbit strength: the columns correspond to $\omega=0.1, 0.5, 1.5$ while the rows correspond to $J_\alpha=0,0.01,0.1$. The current is applied in \textit{opposite direction} as the spin-wave driven domain-wall motion.}
\label{fig:with2} 
\end{figure*}

\begin{figure}[t!]
\centering
\resizebox{0.49\textwidth}{!}{
\includegraphics{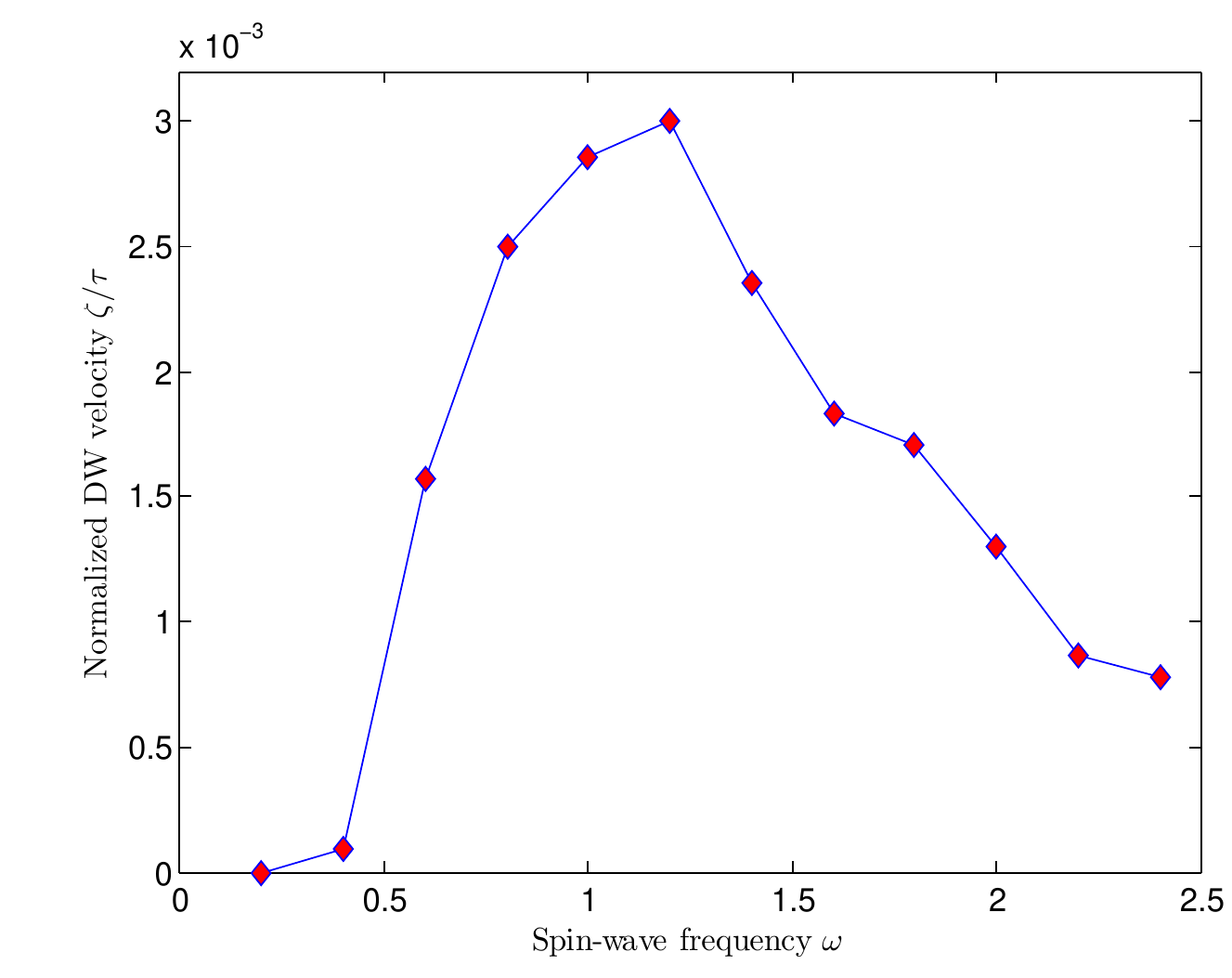}}
\caption{(Color online) Frequency-dependence of the domain-wall velocity induced by magnons. We have used the dimensionless parameters $\alpha=0.01, \beta=0.005$, $A=8\times10^{-5}$ with $K_x=-0.5$, $\mathcal{H}_\text{ext,0}=2$ and $|J|=|J_\alpha|=0$ (absence of spin-polarized current.}
\label{fig:velocity} 
\end{figure}

In order to understand the underlying physical mechanism for the termination of the domain-wall motion, it is instructive to consider a simplified analytical model for the time-evolution of the spin-waves. Following a similar prescription as in Ref. \cite{yan_prl_11}, we model the generated spin-wave as a small fluctuation around the original domain-wall configuration: $\mathbf{m} = \mathbf{m}_0 + m_\theta\e{-\i\omega \tau}\hat{\theta} + m_\phi \e{-\i\omega \tau}\hat{\phi}$, where the angular unit vectors are defined relative $\mathbf{m}_0$. Inserting the standard Walker-ansatz with $\tan(\theta_0/2) = \e{x/\Delta}$ while keeping the exchange stiffness and anisotropy terms, we similarly to Ref. \cite{yan_prl_11} find that the solution for the spin-wave may be written as $m_\theta-\i m_\phi = n_0\frac{\i q - \tanh(x/\Delta)}{1+\i q} \e{\i q x/\Delta}$
where $n_0$ is the spin-wave magnitude and $q^2=\omega/K-1$ with $K$ being the anisotropy constant. In the presence of damping, $q$ becomes imaginary. Although this expression no longer strictly holds when including additional terms, such as the spin-orbit interaction, it demonstrates the fundamental mechanism of the magnon spin-transfer torque. A wave incoming from one side $(x\to-\infty)$ reverses its spin as it passes through the domain-wall $(x\to\infty)$, thus depositing a torque of spin angular momentum which causes the domain-wall to propagate in the opposite direction. 

Based on this observation, we may now explain why the domain-wall motion is arrested. In Fig. \ref{fig:without}, the spin-waves initially deposit their spin torque to the domain-wall when approaching it from the left side, causing it to move in the same direction. However, when the domain-wall has travelled far enough to pass the point where the spin-waves are generated, the spin-waves now impinge on the domain-wall center from the right side and thus causes it to move in the \textit{opposite} direction. As long as the magnon torque exceeds the current-induced torque, this suggests that the domain-wall will be pinned in a specific location.

We now consider in more detail how the magnetization dynamics, and in particular the reversal and termination of the domain-wall motion, depends on the field-frequency $\omega$ and the spin-orbit coupling strength $J_\alpha$. This is done by including a finite value of $J_\alpha$ in Eq. (\ref{eq:llg}), which models the influence of the effective field induced by spin-orbit coupling. Since this field is proportional to the applied current, we require that $\text{sign}\{J_\alpha\}$ = $\text{sign}\{J\}$, and note that the spin-orbit field changes direction along with the current.  As with the other current-induced terms, there is an adiabatic and non-adiabatic contribution to this term, the latter proportional to $\beta$. In Fig. \ref{fig:with1}, a spatiotemporal map of the domain-wall motion is provided for several values of $\omega$ and $J_\alpha$ in the case where the current $J$ is applied in the same direction as the magnon induced domain-wall motion. For small values of $\omega$, the magnons have only a minor effect on the domain-wall velocity whereas with increasing $\omega$, the domain-wall is accelerated to begin with. This is reasonable as the magnon-generation should be less efficient as the field becomes more constant, $\omega\to 0$. However, the change in dynamics as the frequency is increased is more complicated than merely an increase in the domain-wall velocity. In fact, there seems to be an intermediate region around $\omega\simeq 0.5$ where the domain-wall structure is less clearly defined. This indicates that there may exist a resonance frequency regime where the domain-wall profile is more easily modified. We plot in Fig. \ref{fig:velocity} the frequency-dependence of the domain-wall velocity induced by the magnons in the absence of any spin-polarized current. As seen, there is an optimal frequency-regime where the induced velocity is at its maximum. Such a non-montonic dependence on the field-frequency $\omega$ is consistent with the observations in Ref. \cite{yan_prl_11}. 

Turning to the role of the spin-orbit interactions, there are two interesting features emerging from Fig. \ref{fig:with1}. Firstly, it is seen that the general trend of the spin-orbit coupling is to destabilize the domain-wall profile, inducing an oscillating magnetization pattern. Secondly, the termination effect of the domain-wall motion is cancelled for sufficiently strong spin-orbit coupling $(J_\alpha\leq0.1$). 
To understand these features, one may note that the influence of spin-orbit coupling in the effective field is seen to mainly amount to a current-dependent field in the transverse direction of the wire. The transverse spin-orbit field eventually destabilizes the domain-wall structure after the application of a current. The magnetization dynamics shown in Fig. \ref{fig:with1} suggest that the influence of the magnon-induced torque is suppressed as the spin-orbit coupling is increased. 

The same tendency is found with regard to the domain-wall reversal shown in Fig. \ref{fig:with2}, where the current is applied in the opposite direction. The frequency-dependence is similar to Fig. \ref{fig:with1}: the magnon-induced torque becomes smaller with decreasing $\omega$ and there is no reversal below a threshold value $\omega_c$. In addition to the destabilized domain structure induced by increasing spin-orbit coupling, it is seen that the domain-wall reversal also becomes less efficient with increasing $J_\alpha$. This supports the conclusion that the spin-orbit interaction suppresses the influence of the magnons in the system. Although an analytical description of the precise frequency-dependence is not possible to achieve, it is clear from the numerical results that the domain-wall reversal and termination are only stable as long as the spin-orbit interaction is sufficiently small (we find that the qualitative behavior in Fig. \ref{fig:without} persists up to around $|J_\alpha|\sim0.01$).

\section{Summary}

In conclusion, we have investigated the interplay between the torque exerted by a spin-polarized current and magnons on a domain-wall present in a ferromagnetic wire. Using a realistic set of parameters, we have demonstrated that the magnon-induced torque is capable of reversing the domain-wall motion in an oscillatory fashion. Moreover, it has been shown that the domain-wall motion may be be terminated and pinned at specific locations determined by the origin of the spin-waves. We have also considered how the presence of spin-orbit interactions influences the above results. These phenomena might open new perspectives with regard to control and manipulation of domain-wall motion as compared to the conventional mean of current-induced torque.


\begin{thebibliography}{999}
\bibitem{para} Our dimensionless parameters correspond to the real unit values $M_s$ = 5$\times10^5$ A/m, $H_x=0.4$ T, $A_\text{ex} = 10^{-11}$ J/m, $P=0.7$, $L=2$ $\mu$m, $j_e=3\times10^6$ A/cm$^2$, and $\omega = 80$ GHz, which correspond well with an experimentally relevant scenario \cite{kim_arxiv_11}.
\end{thebibliography}
\end{document}